\documentstyle[epsf,twocolumn]{jpsj}
\catcode`\@=11
\def\runauthor{Masatoshi Imada and Tsuyoshi Kashima}
\def\runtitle{Path Integral Renormalization Group Method  for Strongly Correlated Quantum 
Systems}
\def\eqlt{\mathrel{\mathpalette\@vereq<}}  % < over =
\def\eqgt{\mathrel{\mathpalette\@vereq>}}  % > over =
\def\@vereq#1#2{\lower2.5pt\vbox{\baselineskip0pt \lineskip-.5pt
  \ialign{$\m@th#1\hfil##\hfil$\crcr#2\crcr{=}\crcr}}}

\def\simlt{\mathrel{\mathpalette\@vereq<}}  % < over \sim
\def\simgt{\mathrel{\mathpalette\@vereq>}}  % > over \sim
\def\@versim#1#2{\lower2.5pt\vbox{\baselineskip0pt \lineskip-.5pt
  \ialign{$\m@th#1\hfil##\hfil$\crcr#2\crcr{\sim}\crcr}}}

\title{Path-Integral Renormalization Group Method for Numerical Study of
Strongly 
Correlated Electron Systems\footnote{submitted to J. Phys. Soc. Jpn}}
\author{Masatoshi Imada and Tsuyoshi Kashima}
\inst{ Institute for Solid State Physics, University of Tokyo, Kashiwanoha,
Kashiwa, 277-8581}
\recdate{ June 13, 2000 }
\abst{A numerical algorithm for studying strongly correlated electron systems is proposed. The groundstate wavefunction is projected out after numerical renormalization procedure in the path integral formalism.  The wavefunction is expressed from the optimized linear combination of retained states in the truncated Hilbert space with numerically chosen basis.  This algorithm does not suffer from the negative sign problem and can be applied to any type of Hamiltonian in any dimension.  The efficiency is tested in examples of the Hubbard model where the basis of Slater determinants is numerically optimized. We show results on fast convergence and accuracy achieved with small number of retained states.  }

\kword{quantum simulation, strongly correlated electrons, Hubbard model, numerical renormalization group}

\begin{document}
\sloppy
\maketitle

%\section{Introduction} \label{SECTION_intro}

Numerical methods for studying interacting quantum particles have been explored extensively in these decades under the rapid progress of computer power.
In particular, correlated electron systems have been the subject of continuing intensive studies.  Useful and efficient algorithms for interacting fermions  contribute to understanding of  many fundamental issues in condensed matter physics as well as to getting insight and prescriptions for future potential applications of electronics.
However, since the proposal of Feynman~\cite{Feynman}, simulation methods  of interacting fermions have remained challenging due to several reasons.

Quantum Monte Carlo (QMC) method is one of the powerful techniques for correlated quantum systems and it has indeed made possible to clarify various aspects of strong electron correlations near the Mott insulator in two dimensions~\cite{Imada1,Imada2,Imada3}.  
 However, for further efficient simulations,  the QMC method on
fermion systems is known to suffer from the negative sign
problem~\cite{FurukawaMinussign}. If the sign problem arises after
sampling or truncation of the full Hilbert space, it  leads to exponentially damped signals for increased system sizes and lowered temperatures under the cancellation of positive and negative terms, thereby makes reliable estimate practically impossible in the presence of statistical or round-off  errors.  

Density matrix renormalization group developed by White~\cite{White}
is another technique for fermion systems.  It offers an efficient way
to reach the ground state wavefunction as well as thermodynamic
quantities without the sign problem.  However, this method can be
applied efficiently only to systems with one-dimensional configurations 
under open boundary conditions.  

In this letter, we propose a new alternative algorithm to compute the
ground state properties.  In our method, the optimized  ground-state
wavefunction $\vert \Phi \rangle$ is obtained as a linear combination
of states as $\vert \Phi \rangle = \sum_l c_l \vert \varphi_l \rangle$
within the allowed dimension, $L$, of the Hilbert space in a numerically chosen basis $\{\vert \varphi_l \rangle\}$.  The ground state is projected out after successive renormalization processes in the path integral.  Our renormalization method optimizes both the basis $\vert \varphi_l \rangle$ and the coefficients $c_l$.  From its formalism, it is apparent that this method is completely free from the sign problem, because the explicit form of the ground state wavefunction is constructed.  Physical quantity $A$ can be calculated from $\langle A \rangle = \langle \Phi \vert A \vert \Phi \rangle / \langle \Phi \vert \Phi \rangle$.
This may be viewed as a numerical procedure to find the best
variational ground-state wavefunction within the allowed dimension of
the Hilbert space, $L$.  With increasing $L$, $\vert \Psi\rangle$ can be
systematically improved from the starting variational state at $L=1$
chosen as such as the Hartree-Fock and RVB states.

%\section{Algorithm} \label{ALGORITHM}

In principle, the renormalization is achieved by successively
operating the projection operator $\exp[-\tau H]$ with a finite $\tau$
 or $(H+c)$ with an appropriate constant $c$ to the initial trial wavefunction $\vert \Phi_0 \rangle$.  The
operations project out higher energy states and the lower and lower
energy states are obtained.  It is viewed as the numerical
renormalization in the imaginary time direction.  
After operating $\exp[-\tau H]$ or $(H+c)$, the dimension of the obtained state in our chosen basis of the Hilbert space expands through the off-diagonal element of $H$ in the given basis representation.  In the process of successive operations, the dimension exponentially increases.  
If the system size is small enough, the whole Hilbert space can be stored in the computer memory.  However, with increasing system size, of course, the whole space cannot be stored and we seek for the best truncation of the Hilbert space within the allowed memory and computation time. Under the constraint that the number of the stored states, $L$, is kept constant,  we repeat the process of the projection and the truncation to lower the energy of the resultant wavefunction.  

For the renormalization, we repeatedly operate $\exp [-\tau H]$ with small $\tau$ or $(H+c)$ to obtain the ground state as  $  \vert \Phi \rangle = \exp [-\tau H]^p \vert \Phi_0 \rangle$ or $  \vert \Phi \rangle = (H+c)^p \vert \Phi_0 \rangle$ for large $p$.  In this paper, we take the first choice $\exp [-\tau H]$ for the projection. The iterative renormalization process is performed in the following steps.  When $\vert \Phi_p \rangle = \sum_{l=1}^L c_l^{(p)}\vert  \varphi_l^{(p)} \rangle$ is given from the previous ($p-1$)-st step,  $\vert \Psi_{p} \rangle=\sum_{l=1}^{L}\sum_{j=1}^J c_l^{(p)}\vert  \psi_{l,j}^{(p+1)} \rangle $ in the $p$-th step is computed where the set $\psi_{l,j}^{(p+1)}$ is provided from $\sum_{j=1}^J \vert  \psi_{l,j}^{(p+1)} \rangle = \exp [-\tau H]\vert \varphi_l^{(p)} \rangle$ and $j$ denotes each term in the summation over the space expanded through the operation $\exp [-\tau H]$ in the low order of $\tau$.  Thus the dimension of the Hilbert space is expanded due to the summation over $j$.  To keep the dimension at the $p$th step, we next pick up $L$ states, $\{\varphi_l^{(p+1)}\}$ out of  the expanded states $\psi_{l,j}^{(p+1)}$. To select $L$ states out of $LJ$ states in $\psi$, we solve a generalized eigenvalue problem 
\begin{eqnarray}
\sum_n H_{m,n}c_n=\lambda \sum_n F_{m,n}c_n,
\label{1}
\end{eqnarray}  
 where $H_{m,n}=\langle 
\varphi_m^{(p+1)}\vert H \vert \varphi_n^{(p+1)} \rangle$ and  $F_{m,n}=\langle 
\varphi_m^{(p+1)}\vert \varphi_n^{(p+1)} \rangle$.   Note that the
basis functions $\vert \varphi_i \rangle$ generated after the
operation of $\exp [-\tau H]$ is not necessarily orthogonalized.  If
so, the eigenvalue problem has to be generalized accordingly due to nonzero $F_{m,n}$ for $m\not= n$. For each candidate of the truncated set $\{\vert\varphi_m^{(p+1)}\rangle\}$, we calculate the lowest eigenvalue $\lambda_0$ and compare them.  The set $\vert\varphi_l^{(p+1)}\rangle$ ($l=1...,L$) are employed when it gives the lowest $\lambda_0$ among the candidates, $\vert\psi_{l,j}^{(p+1)}\rangle$.  The coefficients $c_n^{(p+1)}$ are automatically given from the eigenvector with the lowest eigenvalue $\lambda_0$ in the above generalized eigenvalue problem (\ref{1}).  
Then the  $(p+1)$-st renormalized state is given as $\vert \Phi_{p+1} \rangle = \sum_l c_l^{(p+1)}\vert  \varphi_l^{(p+1)} \rangle$.

There may exist several ways to pick up $L$ states out of $LJ$
states. When the whole  $\exp [-\tau H]$ is operated, the
expansion $J$ in general becomes huge for large system sizes, because each  local
term of the Hamiltonian may generate off-diagonal terms.  Then the number of 
truncating ways becomes exponentially large.  To avoid it, several ways are possible.  One possible way is to rewrite $\exp [-\tau H]$ as $\prod_i \exp [-\tau H_i]$ with local terms $H_i$, and perform the truncation process after each operation of the local term $\exp [-\tau H_i]$.  We call this the local algorithm. Another way is to introduce a stochastic sampling, where, among huge number of states expanded after the operation of $\exp [-\tau H]$, sampling and annealing are performed  using the Monte Carlo method and one sample is employed after annealing.  We call it the Monte Carlo truncation algorithm.   

To present our algorithm in a more concrete way, we take the Hubbard model defined by  
\begin{subeqnarray}
{\cal H} & = &{\cal H}_t + \sum_i H_{Ui} -\mu M  \slabel{3.1} \\
{\cal H}_t & = & -t\sum_{\langle
ij\rangle}(c^{\dagger}_{i\sigma}c_{j\sigma}+h.c.) \slabel{3.2} \\
\mbox{and}\quad \quad \quad \quad \quad &\  &\nonumber \\
{\cal H}_{Ui} & = & U (n_{i\uparrow}-\frac{1}{2})(n_{i\downarrow}-\frac{1}{2})\slabel{3.4},
\end{subeqnarray}  
where 
$M  \equiv  \sum_{i\sigma} n_{i\sigma}$.  
Here, we follow the standard notation.   In our example of the Hubbard model, we employ, as the chosen basis, 
Slater determinants, $\vert  \varphi \rangle$ whose matrix representation has $N$ by $M$ matrix for $M$ fermions in an $N$-site system.  We first prepare linearly independent $L$ Slater determinants $\vert  \varphi_l^{(0)} \rangle$ ($l=1,...L$) and their coefficients $c_l^{(0)}$ to start with the initial state $\vert \Phi_0 \rangle = \sum_l c_l^{(0)}\vert  \varphi_l^{(0)} \rangle$. 
To realize faster convergence¡¤it is useful to take, as the initial
trial $\vert  \varphi_l^{(0)}\rangle$, for example, a set of samples generated in the conventional quantum Monte Carlo calculations.  

We obtain the optimized wavefunction, $\vert \Phi \rangle = \sum_l
c_l\vert  \varphi_l \rangle$, after sufficient number of
renormalization process applied to this initial wavefunction.  We note 
that the best Hartree-Fock result should be reproduced at $L=1$ and
increasing $L$ systematically improves it.  

To operate $\exp [-\tau H]$ to a Slater determinant,
we first take the path integral formalism and $\exp [-\tau H]$ is approximated by $\exp [-\tau (H_t- \mu M)]\exp [-\tau \sum_iH_{Ui}]$ for sufficiently small $\tau$.  Then we use the Stratonovich-Hubbard transformation for the interaction part.  The interaction part $\exp[-\tau H_{Ui}]$ is replaced with the sum over the Stratonovich variable $s$ as 
\begin{eqnarray}
\exp [-\tau H_{Ui}] & = & \frac{1}{2}\sum_{s=\pm 1}
\exp[2as(n_{i\uparrow}-n_{i\downarrow})\nonumber \\
& - & \frac{U\tau}{2}(n_{i\uparrow}+n_{i\downarrow})]
\label{3}
\end{eqnarray}  
where $a=\tanh^{-1}\sqrt{\tanh(\frac{U\tau}{4})}]$.
Because of the summation over the Stratonovich variables, the number of states to be kept after the operations of $\exp [-\tau H]$ exponentially increases with increasing number of operations.

Then truncation of the states is performed following the above mentioned procedure.  
When the kinetic term $ \exp[-\tau(H_t - \mu M)] $ is operated, the dimension in the Hilbert space does not increase.  The dimension increases from $L$ to $L+2$ when the local interaction term $\exp[-\tau H_{Ui}]$ is operated to $\vert \varphi_l^{(p)}\rangle$.  
Here, we newly add two states obtained from the operation of Eq.(3) to $\vert \varphi_l^{(p)}\rangle$ in addition to the original $\vert\varphi_l^{(p)}\rangle$ itself.  This increases the whole number of states from $L$ to $L+2$.  The original $\vert\varphi_l^{(p)}\rangle$ is also retained as a candidate simply to reach a better estimate.
In the local algorithm, the truncation from $L+2$ to $L$ is taken by solving the generalized eigenvalue problem every time at each operation of $\exp[-\tau H_{Ui}]$ to $\vert \varphi_l^{(p)}\rangle$ and by finding the eigenstates with the lowest eigenvalue among the sets of $L$ retained states. This projection and truncation processes are repeated $NL$ times to complete a unit operation of $\exp[-\tau H]$ to the $L$ retained states.  
In the Monte Carlo truncation algorithm, $\prod_i \exp[-\tau H_{Ui}]$ is applied to $\vert \varphi_l^{(p)}\rangle$, which generates new $2^N$ terms.  If this projection is performed from $p=1$ to $p=L$, it generates new $2^NL$ states. In this case, $2^N$ corresponds to $J$ defined before. For truncation, one sample with $L$ states is employed among $2^NL$ states after Monte Carlo sweeps of the Stratonovich variables to pick up lower energy states.  The convergence to the optimized states seems to become faster when the above two procedures are combined and taken alternately.  Each stored state $\vert \varphi_l\rangle$ can be viewed as analog of each sample in the QMC method.  However, an important difference is that the present method is free of the sign problem because the stored $L$ states are used to generate a many-body wavefunction by the optimized linear combination of them.  Another difference is that the coefficients of the states,$c_l$, in the linear combination are also optimized for a better approximation of the ground state in contrast with the QMC method where the weight of each sample is taken unity in the importance sampling algorithm.

The ground state is approximated by a truncation of the Hilbert space
also in the DMRG method.  In the DMRG method, the truncated basis is
represented after many linear transformations of the original
basis. Although this transformation makes it possible to reach highly
accurate results, the transformed basis is not represented easily
from a simple basis we usually take.   This is the reason why
the application of DMRG is restricted to the cases with the
one-dimensional configuration.  Otherwise  the operator of the physical
quantities are not easily represented in the transformed basis.   In contrast with
DMRG, our algorithm allows basis only when the matrix elements can be easily 
estimated.  This constraint would make the truncation error large for fixed $L$, while it
allows calculations in any type of lattice structures and boundary
conditions.    
Truncations of the Hilbert space were also examined in several other proposals~\cite{deRaedt,Prelovsek,Riera,Honma,Modine}. However, to reach faster convergence at smaller $L$, it is crucial to carefully optimize the wavefunction iteratively within a fixed $L$ as in the procedure of the present work.

We summarize our renormalization process in the following.

\noindent
(1) :Initial state 

\noindent
A set of truncated basis $\vert \varphi_l^{(0)} \rangle$ ($l=1,L$)  and the coefficients 
$c_l^{(0)}$ are given to generate the initial variational wavefunction $\vert \Phi_0  \rangle= 
\sum_l^L c_l^{(0)}\vert \varphi_l^{(0)} \rangle$ .

\noindent
(2): Projection 

\noindent
$\vert \varphi_l^{(0)} \rangle$ are evolved by the operation of $\exp [-\tau H]$.
After this evolution, the number of states to be kept is increased and thus the dimension in the Hilbert space is expanded.  

\noindent
(3): Truncation

From the increased states, we select the same number of states, $L$, as the beginning and discard the rest to keep the dimension in the Hilberd space.  This selection is numericaly performed to realize the lowest energy states. In the numerical process, such lowest energy state is searched after the numerical optimization of the coefficients $c_i$ by examining proposed ways of the truncations.  In each possible way of the truncations, $r$, the lowest eigenvalue $\lambda_0$ and eigenvector $c_l$ for a generalized eigenvalue problem $\langle 
\Phi_r\vert H \vert \Phi_r \rangle = \lambda \langle \Phi_r\vert \Phi_r \rangle$ is solved and $\lambda_0$'s are compared to select the lowest energy state.

The steps (2) and (3) constitute a unit renormalization procedure. 

\noindent
(4): Iteration

The chosen sets $\vert \varphi_l^{(1)} \rangle$ and $c_l^{(1)}$ ($l=1,...,L$) are used as the starting point of the next-step renormalization procedure.

\noindent
(5): Convergence

This renormalization process is repeated until the energy converges to the lowest energy under the allowed dimension of the Hilbert space.

Since we are able to treat only a subspace of the whole Hilbert space in large systems, it is important to analyze the extrapolation procedure using the calculated results under constrained Hilbert space.  At the moment, it is not clear enough how physical quantities converges to the exact value as a function of $L$.  However, it can be shown that the quantities linearly converge as a function of the energy variance defined by 
$\Delta_E = (\langle E^2 \rangle -\langle E \rangle^2)/ \langle E
\rangle^2$
 if $\Delta_{E}$ is small~\cite{Sorella}, where 
$\langle E^2 \rangle=\langle \Phi_r\vert H^2 \vert \Phi_r \rangle / \langle \Phi_r\vert \Phi_r \rangle$ and 
$\langle E \rangle= \langle \Phi_r\vert H \vert \Phi_r \rangle/
\langle \Phi_r\vert \Phi_r \rangle$.  Note that the variance vanishes
if the exact ground state is obtained.  The convergent 
result in the ground state  may be obtained after extrapolation to the zero variance,
although we have to be careful in excluding possible accidental
convergence to an excited eigenstate with increasing $L$.

\begin{figure}[hbt]
%\begin{center}\leavevmode
\epsfxsize=7.6cm
$$\epsffile{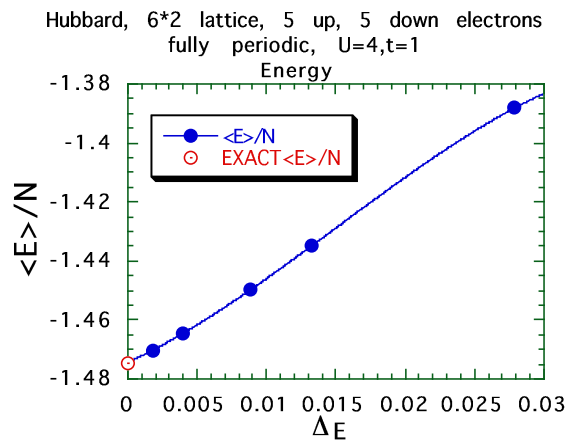}$$
\caption{The calculated energy of the Hubbard model at $t=1$ and $U=4$ as a function of the energy variance $\Delta_E$ for 5 up and 5 down electrons on the $6 \times 2 $ lattice with the periodic boundary condition.}
\label{Fig1}
%\end{center}
\end{figure}

Here we show a few examples for the Hubbard model to show the
efficiency of this numerical procedure.  We show
results at $U/t=4$. 
The first example is the case of 5 up and 5 down electrons on the 6 by
2 lattice with the full periodic boundary condition.  The convergence
of the ground state energy to the exact value is plotted in Fig.1,
where the exact value is reproduced in 4 digits after the extrapolation
 $\Delta_E \rightarrow 0$.  The obtained energy per site is -1.4744
and the exact value is -1.4746.  The largest $L$ we used in this extrapolation is 400.
The momentum distribution $<n_k>=<c_k^{\dagger}c_k>$ and the spin structure factor $S(q)$ obtained after the extrapolation are compared with the exact values in Table I.

\begin{table}
\caption{Comparison of the equal-time spin correlations $S(q)$ and the 
momentum distribution $n(q)$ between the present results and the exact 
ones for the Hubbard model at $U/t=4$ on a 2$\times$ 6 lattice with
the periodic boundary condition.}

Equal time spin correlation S(k$_{\rm x}$,k$_{\rm y}$)

\vspace{2mm}

\begin{tabular}{lll}
(kx,ky) & \multicolumn{1}{c}{present result} & \multicolumn{1}{c}{exact}\\
(1,0) & 0.03662 & 0.03610\\
(2,0) & 0.04125 & 0.04075\\
(3,0) & 0.04161 & 0.04160\\
(0,1) & 0.1293 & 0.1299\\
(1,1) & 0.1302 & 0.1306\\
(2,1) & 0.1339 & 0.1335\\
(3,1) & 0.1375 & 0.1335
\end{tabular}
%\end{table}

\vspace{5mm}
%\begin{table}
Momentum distribution n(k$_{\rm x}$,k$_{\rm y}$)

\vspace{2mm}
\begin{tabular}{lll}
(kx,ky) & \multicolumn{1}{c}{present result} & \multicolumn{1}{c}{exact}\\
(0,0) & 0.9695 & 0.9681\\
(1,0) & 0.9620 & 0.9601\\
(2,0) & 0.9305 & 0.9281\\
(3,0) & 0.01676 & 0.01850\\
(0,1) & 0.06892 & 0.06806\\
(1,1) & 0.04461 & 0.04513\\
(2,1) & 0.02672 & 0.02787\\
(3,1) & 0.02233 & 0.02281
\end{tabular}
\end{table}

The second example is the case of 13 up and down electrons each on the
6 by 6 lattice with the periodic boundary condition.  The results
obtained from $L$ up to 400 again are compared with the QMC results.
The number of states to be retained seems gradually to increase with the
increasing system size if one wishes to reach the same accuracy.
However, the size scaling for extrapolation to $L \rightarrow \infty$
appears to equally work.  The obtained energy per site, -1.1751 is
compared with the QMC result, -1.1756$\pm$0.002~\cite{Imada2}.   

Here, we note several aspects on efficiency of this algorithm clarified in our study and also discuss problems to be solved in the future.
In the present algorithm, solving the generalized eigenvalue problem (\ref{1}) requires the computation time proportional to $L^3$.  This is the time needed to update one of $\vert \varphi_l^{(p)} \rangle$ to $\vert \varphi_l^{(p+1)} \rangle$.  Then the total computation time is proportional to $L^4$ for updating all the $\vert \varphi_l^{(p)} \rangle$ ($l=1,..,L$) once.  
The evolution $\exp [-\tau H]\vert \Phi^{(p)} \rangle$ requires $LN^2$ operations.  To compute all the matrix elements $H_{m,n}$ and $F_{m,n}$, we need time proportional to $L^2N^3$ because computation of each element  $H_{m,n}$ or $F_{m,n}$ takes time $\propto N^3$.  Since the matrix elements can be calculated from the Green function $G_{l,l'}(k,m)=\langle \varphi_l\vert c_k c_m^{\dagger} \vert \varphi_l' \rangle$,
it is still proportional to $L^2N^3$ in updating $\prod_i \exp[-\tau H_{Ui}]\vert \varphi_l\rangle$ by the local algorithm, because $\langle 
\varphi_l\vert\exp[-\tau H_{Ui}] c_k c_m^{\dagger} \exp[-\tau H_{Ui}]\vert \varphi_l' \rangle$ can be calculated from the old Green function $G_{l,l'}(k,m)$ using the local updating procedure used in the QMC method~\cite{Imada1} with the computation time $\propto N^2$.  
When the convergence to a required precission is reached in a fixed number of renormalization steps, the total computation time is dominated by the larger one between $L^4$ and $L^2N^3$.
This computation time may be compared with the time $N_sN^3$ needed for $N_s$ steps of the QMC calculation.  If $L$ can be suppressed less than $N_s$ to reach the same accuracy, the present algorithm may compete or works more efficiently than the QMC method.  
In fact, $L$ can be suppressed to several hundreds in our examples to
reach the accuracy of three digits in energy, which is a typical
accuracy of QMC calculations with the order of 10$^4$ steps.  In fact, the statistical error in QMC is scaled by $1/\sqrt{N_s}$.  The
advantage of this algorithm is that it does not suffer from the
statistical sampling error as in the QMC method.  Furthermore, the present algorithm is the only efficient way in multi-dimensional systems  if the  sign problem is serious in the QMC calculations.

We next discuss interesting future subject for more 
improved  algorthm.  In our study, a combined algorithm of Monte Carlo
and local ones are tested for the truncation.  To project out
into the ground state, a more efficient method of projection and
truncation could be used.  In fact, if each $\vert
\varphi_l^{(p)}\rangle$ is evolved separately from 1 to $L$, the
projection may become slow possibly due to the multi-minimum structure
of the energy in the Hilbert space.  To overcome it, it would be
useful to implement an efficient algorithm employed in random systems
with many nearly degnerate energy minima structure.

When $L$ becomes large, the most  time-consuming process becomes solving of the generalized eigenvalue problem.  It would be helpful if the computation time of this part would be substantially reduced from $L^4$ operations.  Since the matrix operations have instabilities, a simple method like the conjugate gradient method does not work and a more careful algorithm is needed.

In summary, a new algorithm for computing physical quantities of correlated fermion systems is proposed.  The ground state wavefunction is obtained as a linear combination of states  within the allowed dimension of the Hilbert space in numerically optimized basis.  The optimization is achieved after the numerical renormalization procedure in the path integral form.  The results are extrapolated by systematically examining  the dependence on the dimension of the Hilbert space.  Physical quantities show good convergence and accuracy in the calculation of the Hubbard model.  Since this method does not suffer from the negative sign problem and can treat any type of lattice structure, it opens a possibility for efficient simulations which cannot be reached by existing algorithm such as the quantum Monte Carlo method or the density matrix renormalization group.
%\end{\item}

\acknowledgement
We would like to thank N. Furukawa for useful discussion.
The work is supported by "Research for the Future" program from
the Japan Society for the Promotion of Science under the grant number
JSPS-RFTF97P01103.

%\begin{figure}[hbt]
%\begin{center}\leavevmode
%\includegraphics[width=0.9\linewidth]{6by2} 
%\caption{The calculated energy of the Hubbard model at $t=1$ and $U=4$ as a %function of the energy variance $\Delta_E$ for 13 up and 13 down electrons on %the $6 \times 6 $ lattice with the periodic boundary condition.}
%\label{Fig2}
%\end{center}
%\end{figure}

\end{document}